\documentclass[preprint,prb,tightenlines]{revtex4}
\usepackage{bm}
\begin{document}
\title{Spin-Hall effect in semiconductor heterostructures with
cubic Rashba spin-orbit interaction}
\author{O. Bleibaum}
\email{olaf.bleibaum@physik.uni-magdeburg.de}
\affiliation{Institute for Theoretical Physics, Otto-von-Guericke University,
PF 4120, 39016 Magdeburg, Germany}
\author{S. Wachsmuth}
\affiliation{Institute for Theoretical Physics, Otto-von-Guericke University,
PF 4120, 39016 Magdeburg, Germany}
\begin{abstract}
We study the spin-Hall effect in systems with weak cubic Rashba spin-orbit 
interaction. To this end we derive particle and spin diffusion equations
and explicit expressions for the spin-current tensor in the diffusive 
regime. We discuss the impact of electric fields on the
Green's functions and the diffusion equations and establish a
relationship between the spin-current tensor and the spin diffusion equations
for the model under scrutiny. We use our results to calculate the 
edge spin-accumulation in a spin-Hall experiment and show that there
is also a new Hall-like contribution to the electric current in
such systems.  
\end{abstract}
\maketitle
\section{Introduction}
At present there is much interest in investigations of
spin-charge coupling effects in non-magnetic semiconductor 
heterostructures with intrinsic spin-orbit interaction. 
Particular attention is paid to effects, which permit the 
electric generation of non-equilibrium magnetizations in such systems, 
like the spin-Hall effect (see, e.g. Ref.[\onlinecite{Sinova}]-
[\onlinecite{Wu}])
or the spin accumulation induced 
by an external electric field (see, e.g., Ref. [\onlinecite{Kato}]-
[\onlinecite{Kleinert2}]). In the spin -Hall effect a non-homogeneous
magnetization is created at the sample boundaries in the presence of
electric fields. The spin accumulation, by contrast, 
is a homogeneous non-equilibrium magnetization, which is induced by a
current in the bulk.

While the mechanisms leading to spin accumulations in heterostructures
with intrinsic spin-orbit interaction have  been understood for decades 
the spin-Hall effect is still a rather controversial issue. The first
papers on spin-Hall effects in heterostructures with intrinsic spin-orbit
interaction have focused on investigations of spin-current tensors 
for electron gases with linear Rashba spin-orbit interaction\cite{Sinova}
and for hole gases described by the Luttinger model
\cite{Murakami}. The spin-current tensors were defined as the
expection value of the operator (${\hat v}_i$- velocity operator,
$S_k$ -spin operator, $\{..,..\}$ -anticommutator)
\begin{equation}\label{I1}
{\hat j}_{ik}=\frac{1}{2}\{{\hat v_i},S_k\}.
\end{equation}
The investigations have shown that an electric field 
leads to distortions of the spectrum, which render the spin-Hall 
current of clean systems finite, even if their Hamiltonian does not
explicitly contain $S_z$ ($S_z$-spin-operator for particles 
with spin in the perpendicular direction). This fact has led the authors
of the papers in question to the conclusion that there is a non-vanishing
spin-Hall effect of considerable magnitude in such systems. However,
none of them has been able to establish a connection between the spin-current
tensor and the non-equilibrium magnetization induced at the boundary
of a sample. The silent assumption was  the spin current tensor
enters a diffusion like equation, as discussed, e.g., in 
Ref.[\onlinecite{Murakami}].

Further investigations, however,  have shown that the results on the 
spin-current tensor are not robust. In the linear Rashba model
the spin-Hall conductivity vanishes already in the presence of
weak disorder \cite{Inoue,Raimondi,Loss}. In p-doped systems, described
by  Luttinger models, the situation seems to be more favorable. In 
such systems the spin-Hall conductivity is also finite in the presence of 
disorder\cite{Bernevig,Murakami1}. In spite of this fact most investigations
have focused on systems with linear Rashba spin-orbit interaction, 
which are  technically less demanding.

The absence of spin-Hall currents in the linear Rashba model in the 
presence of weak disorder, however, has not remained the only problem casting
doubts on the predictions of spin-Hall effect. More pressing is the
question of the relevance of the approach itself\cite{Rashba,Zhang}.
It has already been shown in Ref.[\onlinecite{Rashba}] that the 
spin-current tensor in the Rashba model is finite, even in equilibrium.
Therefore, it does not characterize
macroscopic transport phenomena. Diffusion equations, which take into
account also spin-charge coupling effects, have been derived in a number 
of papers for systems with weak Rashba interaction
\cite{Burkov,Mishchenko,OB1,OB2,Bauer,OB4,Damker,OB5}. However, they do 
not only
contain the divergence of the spin-current tensor but also additional 
derivative terms. The investigation of these terms leads to the conclusion
that the spin-current 
tensor in the linear Rashba model
is not a quantity, which is continuous at boundaries\cite{OB2},
as stressed also in Ref.[\onlinecite{Zhang}].
In fact,  the solution to the diffusion equations, which is derived under
the assumption that the spin-current tensor vanishes at hard-wall
boundaries, is not physical. It violates time reversal symmetry\cite{OB5}.
Responsible for this fact are just those terms, which are also present 
in equilibrium.  Alternative definitions for the spin-current tensor
have been suggested in a number of papers (see, e.g., Refs.
[\onlinecite{Zhang2,Sugimoto}]) to circumvent these problems. However, 
although the authors have focused on observable quantities they have
failed to 
establish a connection between the non-equilibrium magnetization 
at the boundary
of a sample and their definition. Moreover, their predictions
differ rather strongly from those, which have been derived on the
basis of Eq.(\ref{I1}). The existence of the spin-Hall effect in clean
samples has meanwhile been  
established experimentally
\cite{Wunderlich} in spite of these problems. The predictions of 
Ref.[\onlinecite{Sinova}]
have essentially proven to be correct. The question, however, how to 
describe this effect theoretically, is still remaining.  

Recently, some
investigations have been published, which address the question how
universal the results to the linear Rashba model are
\cite{Nomura,Nomura2,Krotkov}. The results of these investigations 
show that already deviations from the quadratic dispersion relation
of the unperturbed band  spectrum\cite{Krotkov} or deviations
from the linear dependence of the Rashba spin-orbit interaction
\cite{Nomura2,Schliemann} lead to a non-vanishing spin-Hall current
in the bulk even in the presence of disorder. To obtain
the latter result the authors of the Refs.[\onlinecite{Nomura2}] and
[\onlinecite{Schliemann}] use a cubic Rashba model.
The cubic Rashba model, which  describes the heavy holes in III-V 
semiconductor quantum wells, has been introduced in the Refs.
[\onlinecite{Zulicke1,Zulicke2}]. It differs from the linear Rashba
model in that the interaction depends cubically on the momentum.
Thus, the cubic Rashba model offers the opportunity to study 
spin-Hall transport in semiconductor heterostructures with 
intrinsic spin-orbit interaction in a framework, which keeps the simplicity 
of the original Rashba model. Here we use this opportunity
to address the question, how the non-vanishing spin-Hall current
manifests themselves in the spin diffusion equations, how these
equations differ from those in the linear Rashba model, and whether 
the spin-current tensor (\ref{I1}) is the appropriate quantity, 
which has to be considered, or whether alternative definitions
for the spin-current tensor are more appropriate, as claimed, e.g.,  in
the Refs.[\onlinecite{Zhang2,Sugimoto}]. To answer these
questions we derive particle diffusion equations,
spin diffusion equations and explicit expressions for the spin-current 
tensor, discuss
the impact of external fields on the diffusion process
and use
our results to calculate the magnitude of the magnetization for systems
with weak Rashba spin-orbit interaction. Doing so, we establish
an explicit relationship between the spin-current tensor (\ref{I1}), 
the diffusion
equations, and the non-equilibrium magnetization at the boundary
in the presence of an electric field. Thus our results show that the
spin-current tensor (\ref{I1}) determines indeed the physics in 
the cubic Rashba model. We would like to mention that
a paper has recently been published, in which the spin-Hall effect 
has been investigated numerically in the cubic Rashba model\cite{Nomura}.  
The authors of Ref.[\onlinecite{Nomura}] focus on systems
with strong Rashba interaction, in which the energy level splitting at 
the Fermi surface is large compared to the disorder energy. In contrast to 
them we present analytical calculations and focus on systems in
the opposite regime. While the numerical results of 
Ref.[\onlinecite{Nomura}] apply to systems with strong spin-orbit interaction,
in which the energy level splitting at the Fermi energy due to the
spin-orbit interaction is large compared to the disorder energy, 
we focus on systems with weak spin-orbit interaction, in which the
energy level splitting at the Fermi energy is small compared to the 
disorder energy.

\section{Basic equations}
We consider the holes in the lowest heavy-hole subband of a two-dimensional 
hole gas with Rashba spin-orbit interaction. Their motion is described by 
the Hamiltonian\cite{Zulicke1,Zulicke2,Winkler1,Winkler2,Schliemann,Nomura} 
\begin{equation}\label{BE1}
H=\frac{{\bm p}^2}{2m}+i\frac{N}{2}(p_-^3\sigma_+-p_+^3\sigma_-)+V({\bm r}).
\end{equation}
Here ${\bm p}$ is the momentum operator, $m$ is the effective mass, 
$p_{\pm}=p_x\pm ip_y$, $\sigma_{\pm}=\sigma_x\pm i\sigma_y$ and $N$ is 
a constant characterizing the strength of the spin-orbit interaction. The
potential $V({\bm r})$ is a random white noise potential with Gaussian 
statistics and correlation function
\begin{equation}\label{BE2}
\langle V({\bm r})V({\bm r'})\rangle=\frac{\hbar}{2\pi\nu\tau}
\delta({\bm r}-{\bm r'}),
\end{equation}
where $\nu=m/2\pi\hbar^2$ is the density of states per spin and $\tau$ is 
the scattering time. The vectors ${\bm e}_x$ and ${\bm e}_y$ are the unit 
vectors in the two-dimensional plane. ${\bm e}_z$ is a unit vector
perpendicular to the plane. The indices 1,2 and 3 replace  
the indices $x,y$ and $z$ in sums. Double indices have to be 
summed over in all formulas. Sums over Latin indices run from 1-3, sums over 
Greek indices from 0-3.

The Hamiltonian (\ref{BE1}) describes particles 
with single particle energies $E_{\pm}({\bm k})=\epsilon_k\pm\Delta_k$,
where $\epsilon_k=\hbar^2k^2/2m$, $\hbar k$ is the momentum of the 
particles, and $\Delta_k=|N|\hbar^3k^3$ is the energy level splitting 
due to the Rashba spin-orbit interaction. The spectrum is unstable 
at large $k$, since $\Delta_k\propto k^3$. This instability is an artifact,
which results from the fact that the effective Hamilton (\ref{BE1})
is only derived for systems with small k by means of perturbation
theory. Therefore, the Hamiltonian (\ref{BE1}) can only be used for low 
hole densities. From the
practical point of view these are also those typical achieved in 
experiments\cite{Schliemann}. The investigation of transport processes 
on the basis of the Hamiltonian (\ref{BE1}), however, requires  also
the calculation of integrals with infinite upper bound. To render them 
finite we assume that $\Delta_k$ depends only cubically
on k in the region of interest, but increases slower than 
$k^3$ for $k\to \infty$. In this case we can use the same methods
for the investigation of the system, which have also been used for
systems with linear Rashba spin-orbit interaction (see, e.g., 
Refs.[\onlinecite{OB3,Burkov,Mishchenko,OB1,OB2}]). In particular
we can use the self-consistent Born approximation for the description
of the single-particle excitations. The retarded Green's 
function takes the form
\begin{equation}\label{BE3}
G^R({\bm k}|E)=g_+^R({\bm k}|E)g_-^R({\bm k}|E)(g^{-1}_R({\bm k}|E)
\sigma_0+i\frac{N}{2}(p_-^3\sigma_+-p_+^3\sigma_-))
\end{equation}
in this approximation,
where
\begin{equation}\label{BE4}
g^R_{\pm}({\bm k}|E)=\frac{1}{E-E_{\pm}({\bm k})\pm i\hbar/2\tau},
\end{equation}
$\sigma_0={\bm 1}$ is the unit matrix and 
$g_R({\bm k}|E)=g_+^R({\bm k}|E)|_{N=0}$. The advanced Green's function is 
obtained from Eq.(\ref{BE3}) by hermitian conjugation.
\section{Spin diffusion in the absence of external fields}
\subsection{Particle and spin diffusion equations}
To derive particle and spin diffusion equations we use the ladder 
approximation.
The equation of motion for the Fourier-Laplace 
transform of the particle density $n({\bm \kappa}|E,s)= S_0({\bm\kappa}|E,s)$ 
and the Fourier-Laplace transform of the spin density
${\bm S}({\bm \kappa}|E,s)=S_i({\bm\kappa}|E,s){\bm e}_i$ takes the 
form\cite{OB3,Burkov,OB1,OB2}
\begin{equation}\label{SD1}
[\delta_{\eta\gamma}-\frac{\hbar}{4\pi\nu\tau}
\int\frac{d{\bm k}}{(2\pi)^2}\mbox{tr}(\sigma_{\eta}
G^R({\bm k}+\frac{\bm\kappa}{2}|E+
i\hbar s)\sigma_{\gamma}G^A({\bm k}-\frac{\bm\kappa}{2}|E))]
S_{\gamma}({\bm \kappa}|E,s)={S_0}_{\eta}({\bm\kappa}|E)
\end{equation}
in this case.
Here $s$ is the frequency describing the evolution (it
corresponds to a Laplace transformation with respect to time),
tr symbolizes the trace, and ${S_0}_{\eta}({\bm\kappa}|E)$ is the
initial condition. To simplify Eq.(\ref{SD1}) we use the hydrodynamic 
expansion, viz we expand the kernel of Eq.(\ref{SD1}) with respect
to ${\bm \kappa}$, $N$  and $s$ and take into account only contributions,
which are at most of second order in $\kappa$ and of first order in $s$. 
In this case 
we obtain the following diffusion equations for systems with
weak Rashba spin-orbit interaction after inverse Laplace and Fourier
transformation:
\begin{equation}\label{SD2}
\partial_tn-D\Delta n=0,
\end{equation}
\begin{equation}\label{SD3}
\partial_t{\bm S}+{\bm \Omega}\cdot{\bm S}-D\Delta{\bm S}=0.
\end{equation}
Here $D=E\tau/m$ and ${\bm \Omega}$ is a symmetrical tensor of second rank 
with matrix elements ${\bm\Omega}_{ik}=\Omega\delta_{ik}(1+\delta_{i3})$,
where
\begin{equation}\label{SD4}
\Omega=16\frac{N^2m^3}{\hbar^2}E^3\tau.
\end{equation}
In deriving these equations we have ignored terms of the 
order $N^2\nabla_i^2 S_l$, which yield corrections to the
diffusion coefficient. These corrections
are small with respect to the ratio 
$(\Delta_k|_{\epsilon_k=\epsilon_F}/(\hbar/2\tau))^2$ 
and thus negligible in systems with weak Rashba spin-orbit interaction.

It is instructive to compare these equations with those, which
are obtained for systems with linear Rashba spin-orbit interaction
in the same approximation.
If we replace the cubic spin-orbit interaction in Eq.(\ref{BE1}) 
by a linear interaction, according to
\begin{equation}\label{SD5}
i\frac{N}{2}(p_-^3\sigma_+-p_+^3\sigma_-)\to ({\bm N},{\bm\sigma}\times
{\bm p}),
\end{equation}
where ${\bm N}=N{\bm e}_z$,  we obtain the equations
\cite{Burkov,Mishchenko,OB1,OB2,Bauer}
\begin{equation}\label{SD9}
\partial_t n -D\Delta n -\Omega\tau ({\bm\nabla},{\bm N}\times{\bm S})=0.
\end{equation}
\begin{equation}\label{SD10}
\partial_t {\bm S}+{\bm\Omega}\cdot({\bm S}-{\bm S}_A)-D\Delta {\bm S}
-\omega_s({\bm N}\times{\bm\nabla})\times ({\bm S}-{\bm S}_A)=0.
\end{equation}
Here 
\begin{equation}\label{SD11}
{\bm S}_A=-\tau{\bm N}\times{\bm \nabla}n
\end{equation}
is the non-equilibrium magnetization induced by a density gradient
and $\omega_s$ is a transport coefficient proportional to $D$ (Note that,
although we use the character $N$ to characterize the strength of the 
Rashba spin-orbit interaction both for systems with linear and cubic Rashba
interaction this quantity is different in  both systems and possesses 
different dimension. The structure of the tensor ${\bm\Omega}$ in the 
linear model agrees with that in the cubic model. The dependence
of the quantity $\Omega$ on the microscopic details, however, is 
also different in both models.). 

Comparing the Eqs.(\ref{SD2}) and (\ref{SD3}) with the Eqs.(\ref{SD9})
and (\ref{SD10}) we note the following. Whereas the spin-orbit coupling
gives rise to a coupling between spin and charge in the linear model
there is no coupling between spin and charge in the diffusion
equations for cubic systems at all. There is neither a spin accumulation 
(${\bm S}_A=0$) nor a spin-galvanic current. This fact is very
surprising, since we  would have expected that the structure of the 
diffusion equations is little affected by a simple change in the 
power of the momentum in the Rashba spin-orbit interaction. We would like to
mention that we have also checked this conclusion
by calculating the spin 
accumulation in the Kubo-formalism. The results of this calculation are
in line with those obtained from the diffusion equation and therefore
not presented here. They  
also show that the dc-spin accumulation is zero in the linear 
approximation with respect to the electric field for systems with
weak cubic Rashba spin-orbit interaction. 

The diffusion equations (\ref{SD3}) and (\ref{SD10}), however, 
differ not only in the spin-charge coupling terms. Comparing
both equations further we also note that there is no term linear in the
derivatives in the spin-diffusion equation (\ref{SD10}), neither
in the linear approximation with respect to $N$ nor in higher order.
This fact has far reaching consequences. The coefficient $\omega_s$ 
in Eq.(\ref{SD10}) leads to a precession
of the magnetization around an axis, which is parallel to 
${\bm N}\times{\bm F}$,
if an electric field ${\bm F}$ is switched on. This precession can be 
used for the manipulation of
the electron spin (see, e.g., Refs.
[\onlinecite{OB3,Damker,Wu1,Awschalom1,OB2}]). The corresponding effective
Zeeman field is linear in $F$. In the cubic Rashba-model by contrast,
there is no effective Zeeman-field in  linear approximation with 
respect to $F$. 

The huge difference between the structure of the diffusion equations 
raises 
the question why the physics of both systems is drastically different. 
There is no answer
to the question yet. On the technical level, however, the difference
results from the angle averaging. The cubic Rashba-interaction is 
proportional to $p_{\pm}^3=(\hbar k_{\pm})^3$. The calculation of terms of
first order in the derivatives, e.g. of spin-charge coupling
terms, reduces to integrals of the type
\[\int\frac{d{\bm k}}{(2\pi)^2}{\bm k}k_{\pm}^3f(k)=0,\]
which are zero for arbitrary functions $f(k)$.      
\subsection{The spin-current tensor}
The surprising difference between the structure of the spin diffusion 
equations for systems with linear Rashba spin-orbit interaction and
that for systems with cubic Rashba spin-orbit interaction raises
the question, where spin-charge coupling effects manifest themselves.
To investigate this question further we investigate the spin-current
tensor.  The spin-current tensor is defined as the expection value
$J_{ik}({\bm r}|E,s)$ of the operator
\begin{equation}\label{SC1}
{\hat J}_{ik}({\bm r})=\frac{1}{2}\{{\hat j}_{ik},\delta
({\bm {\hat r}}-{\bm r})\}
\end{equation}
where
\begin{equation}\label{SC2}
{\hat j}_{ik}=\frac{1}{2}\{\sigma_i,{\hat v}_k\},
\end{equation}
${\bm{\hat r}}$ is the position operator and 
${\hat v}_k$ is the velocity operator. To calculate the expection value
 we restrict the 
consideration to time scales, which are large compared to the momentum 
relaxation time. In this case we can use the following relationship
between the spin-current tensor and the Green's functions
\cite{Mishchenko,OB5}:
\begin{equation}\label{SC3}
J_{ik}({\bm r}|E,s)=\frac{\hbar}{4\pi\nu\tau}\int d{\bm r}_1d{\bm r}_2
\mbox{tr}(G^A({\bm r}_1-{\bm r}_2|E){\hat J}_{ik}({\bm r})G^R({\bm r}_2
-{\bm r}_1|E)\sigma_{\kappa})S_{\kappa}({\bm r}_1|E,s).
\end{equation}
Here the Green's functions are those in the position representation.
To simplify the integrals we use again the hydrodynamic expansion and 
obtain the equations
\begin{equation}\label{SC4}
J_{ix}=-D\nabla_iS_x,\hspace{1ex}J_{iy}=-D\nabla_iS_y
\end{equation}
and
\begin{equation}\label{SC5}
J_{iz}=-D\nabla_iS_z+\frac{3}{2}\frac{\hbar}{m}\tau\Omega
({\bm e}_z\times{\bm\nabla })_in.
\end{equation}
Eq.(\ref{SC5}) generalizes the existing results\cite{Schliemann} on the 
spin-Hall current to non-homogeneous systems. To compare Eq.(\ref{SC5})
with the result of Ref.[\onlinecite{Schliemann}] for the spin-Hall
current of a homogeneous system in the presence of an electric field
we use the fact that the impact of the field on the spin-Hall current
can be mimed by means of a concentration gradient in the ohmic
approximation\cite{Mishchenko}. To mime the field we use a particle
density of the form
\begin{equation}\label{SC6}
n({\bm r}|E,s)=n_0-2\nu ({\bm F},{\bm r})  
\end{equation}
where ${\bm F}=-e{\cal{\bm E}}$ ($e$ is the electric charge and 
${\cal{\bm E}}$ is the electric field) and $n_0$ is a constant density.
Doing so, we obtain
\begin{equation}\label{SC7}
J_{iz}=-3\Omega\frac{\hbar}{m}\tau\nu({\bm e}_z\times{\bm F})_i.
\end{equation}
This result coincides with that of Ref.[\onlinecite{Schliemann}].
It shows that there is a non-vanishing spin-Hall current in this system.
The crucial question is, whether this fact gives also rise to a
non-vanishing
spin-Hall effect, viz to a non-equilibrium spin accumulation at the 
boundaries of the sample.

The equations (\ref{SC4}) and (\ref{SC5})
differ strongly from those obtained for
systems with linear Rashba-interaction\cite{OB2,OB5}. The spin-current 
tensor for such systems is even non-zero in equilibrium (see, e.g. 
Refs.[\onlinecite{Rashba,Zhang}]). This fact
manifests itself in terms containing the particle density without 
derivatives\cite{OB3,OB5,Kl}. Such terms are absent in systems with 
weak cubic
Rashba-interaction. Therefore, the spin-current tensor describes
deviations from equilibrium in systems with weak cubic Rashba-interaction 
and thus characterizes transport phenomena. In contrast to the spin diffusion 
equation, however, the spin-current tensor couples also spin with charge. 
The coupling is achieved by 
the  second term on the right hand side (rhs) of Eq.(\ref{SC5}).
Since such spin-charge coupling terms are absent in the diffusion 
equation there is the question, whether there is really a connection
between the spin  diffusion equation and  the spin-current
tensor. Such a connection has not been established
for systems with linear Rashba-interaction yet\cite{OB2,OB5}.

From the phenomenological point of view we expect that the
spin diffusion equation can be written in the form\cite{remark}
\begin{equation}\label{SC8a}
\partial_tS_k+{\bm\Omega}_{ki}S_i+\nabla_i\tilde J_{ik}=0.
\end{equation}
Here ${\tilde J}_{ik}$ is a tensor of currents, which we call 
transport currents. 
The transport currents determine the boundary conditions needed to
find solutions to the diffusion equation. The physical solutions are those 
with the property that the transport currents are continuous at the 
boundary in the presence of a boundary\cite{Landau}.
To find  the transport currents there is no other way
than to read off all terms, which are under the sign of the derivative.
However, since Eq.(\ref{SC8a}) contains only the divergence of the
spin-current tensor it is impossible to retrieve terms proportional
to a curl unambiguously.
 
The spin-charge coupling term in Eq.(\ref{SC5}) has just just the 
structure of a curl. Due to this fact the relationship
\begin{equation}\label{SC8}
\nabla_iJ_{ik}=-D\Delta S_k
\end{equation}
holds, so Eq.(\ref{SD3}) can also be written in the form
\begin{equation}\label{SC9}
\partial_tS_k+{\bm\Omega}_{ki}S_i+\nabla_iJ_{ik}=0.
\end{equation}
This fact suggests that the spin-current tensor coincides
with the  tensor of transport currents. Spin-charge coupling effects 
would enter the diffusion equation via the boundary condition if this
assumption were correct.

Close to equilibrium, however,  there is no way to assure 
that  the spin-current tensor really coincides with 
the tensor of transport currents, since the terms proportional to a 
curl can not be retrieved from the diffusion equation, 
as mentioned before. Thus, there is no reason to believe that 
the spin-current tensor really determines the boundary conditions. There is 
no generally accepted method to avoid this problem. However, in a recent 
paper \cite{OB2} we have noted the following fact: we can get further insight 
into the structure of the terms in question by switching  on an additional
external electric field ${\bm F}$. The transport coefficients become
position dependent in this case, since they depend on the kinetic energy 
$\mu_{\bm r}=E-({\bm F},{\bm r})$ of the particles, which is different for 
particles with fixed total energy $E$ at different positions ${\bm r}$. 
Due to this fact the derivatives 
in the diffusion equation have also to act on the transport coefficients
and thus, they produce additional terms. If the assumption that the
spin-current tensor coincides with the tensor of transport currents
is correct, these terms are  just those, which result from derivatives
of the spin-charge coupling term on the rhs of Eq.(\ref{SC5}).
Thus, this observation is helping us to identify unambiguously all those
terms in the tensor of transport currents, which have the structure
\[\omega(\mu_{\bm r})({\bm e}_z\times{\bm \nabla})_iS_{\gamma}({\bm r}|E,s),\]
where $\omega(\mu_{\bm r})$ is an arbitrary transport coefficient.
Only terms of the form
\[({\bm e}_z\times{\bm \nabla})_i(
\omega(\mu_{\bm r})S_{\gamma}({\bm r}|E,s))\]   
can not be retrieved.
\section{Spin diffusion in the presence of an electric field}
\subsection{Diffusion in the presence of an electric field}
To generalize the diffusion equations to the presence of an 
electric field we use the symmetry properties of the Green's functions.
The Green's functions satisfy the property
\begin{equation}\label{F1}
G({\bm r}+{\bm a},{\bm r'}+{\bm a}|E)=G({\bm r},{\bm r'}|E-({\bm F},{\bm a}))
\end{equation}
in the presence of a constant field ${\bm F}$,
where $G({\bm r},{\bm r'}|E)$ is either the retarded or the 
advanced Green's function in the position representation and ${\bm a}$
is an arbitrary vector. This fact permits the introduction of a
new function ${\tilde g}$, according to the relationship\cite{OB1}
\begin{equation}\label{F2}
G({\bm r},{\bm r'}|E)=\int\frac{d{\bm k}}{(2\pi)^2}e^{i({\bm k},{\bm r}-
{\bm r'})}{\tilde g}({\bm k}|\mu_{\bm R}),
\end{equation}
where ${\bm R}=({\bm r}+{\bm r'})/2$. The knowledge of the function 
${\tilde g}$ is sufficient to derive the diffusion equation. 
In the presence of the
field it takes the form\cite{OB1}
\begin{equation}\label{F3}
[a_{\gamma\delta}(\mu_{\bm r},s)+\frac{1}{2}\{{b}^i_{\gamma\delta}
(\mu_{\bm r}),{\nabla}_i\}+
\frac{1}{2}\nabla_i c_{\gamma\delta}^{ij}(\mu_{\bm r})\nabla_j]
S_{\delta}({\bm r}|E,s)={{S_0}_{\gamma}}({\bm r}|E),
\end{equation}
where $\{\dots,\dots\}$ is the anticommutator. The coefficients $a$, 
$b$ and $c$ in this equation are related to  the function ${\tilde g}$
by the relationships
\begin{equation}\label{F4}
a_{\gamma\delta}(\mu_{\bm r},s)=\delta_{\kappa\delta}-
\frac{\hbar}{4\pi\nu\tau}
\int\frac{d{\bm k}}{(2\pi)^2}\mbox{tr}[\sigma_{\gamma}
{\tilde g}^R({\bm k}|\mu_{\bm r}+i\hbar s)\sigma_{\delta}
{\tilde g}^A({\bm k}|\mu_{\bm r})],
\end{equation}
\begin{equation}\label{F5}
b^j_{\gamma\delta}(\mu_{\bm r})=
-\frac{\hbar}{4\pi\nu\tau}
\int\frac{d{\bm k}}{(2\pi)^2}\mbox{tr}[\sigma_{\gamma}
{\tilde g}^R({\bm k}|\mu_{\bm r})\sigma_{\delta}
i\nabla_j{\tilde g}^A({\bm k}|\mu_{\bm r})]
\end{equation}
and
\begin{equation}\label{F6}
c^{ik}_{\gamma\delta}(\mu_{\bm r})=
\frac{\hbar}{4\pi\nu\tau}
\int\frac{d{\bm k}}{(2\pi)^2}\mbox{tr}[\sigma_{\gamma}
{\tilde g}^R({\bm k}|\mu_{\bm r})\sigma_{\delta}
\nabla_i\nabla_k {\tilde g}^A({\bm k}|\mu_{\bm r})].
\end{equation}
The function ${\tilde g}$ can be calculated perturbatively. It depends 
only on $F^2$
in the absence of the spin-orbit interaction and
reduces to
the equilibrium Green's function in such systems in the ohmic approximation. 
The coefficients $a$, $b$ and $c$ are the same in this case as those in the 
absence of the field. Therefore, the impact of electric fields can
be reduced to position dependent shifts of the reference point of the
kinetic energy, which can be taken into account by
a simple substitution of the type 
${\bm \nabla}\to{\bm\nabla}-{\bm F}\partial_{\epsilon}$ 
in the equilibrium
diffusion equation, where $\epsilon$ is the kinetic energy.

However, in the presence of the spin-orbit interaction
this concept proves to be inadequate. The impact of the field on 
transport processes can not be reduced to simple
shifts of the reference point of the kinetic energy in such systems,
as claimed, e.g., in the Refs.[\onlinecite{Mishchenko}] and 
[\onlinecite{Bauer}]. In such systems the field leads both
to a position dependent shift of the reference point of the kinetic
energy and to distortions of the particle spectrum. These
distortions are responsible for the spin-Hall effect, as discussed in
Ref.[\onlinecite{Sinova}]. To take into account the distortions
we restrict the consideration to the ohmic approximation. In this case
we can write the function ${\tilde g}$ in the form
\begin{equation}\label{F7}
{\tilde g}({\bm k}|E)=G({\bm k}|E)+{\tilde g}^{(1)}({\bm k}|E),
\end{equation}
where ${\tilde g}^{(1)}({\bm k}|E)$ is the linear contribution 
to ${\tilde g}$ with respect to $F$ and the function 
$G({\bm k}|E)$ is the equilibrium Green's function. The function
${\tilde g}^{(1)}$ can be calculated perturbatively. For the linear
Rashba-model we obtain 
${\tilde g}^{(1)}\propto N^2({\bm F},{\bm\sigma}\times{\bm k})$
(see Eq.(10) of Ref.[\onlinecite{OB1}]), as expected of the
spin-Hall term. The cubic Rashba-model yields
\begin{equation}\label{F8}
{\tilde g}^{(1)}({\bm k}|E)=i\frac{3\hbar}{2}\sigma_zN^2
(g_+({\bm k}|E)g_-({\bm k}|E))^2(p_-^3p_+^2F_+-p_+^3p_-^2F_-),
\end{equation}
where $F_{\pm}=F_x\pm iF_y$. If we take into account this term in  
calculating  the coefficients in Eq.(\ref{F3})
we obtain the equation
\begin{equation}\label{F9}
\partial_tS_k({\bm r}|E,t)+{\bm\Omega}_{ki}(\mu_{\bm r})S_i({\bm r}|E,t)
+\nabla_i J_{ik}({\bm r}|E,t)=0
\end{equation}
in the ohmic approximation,
where ${\bm \Omega}(\mu_{\bm r})={\bm \Omega}|_{E=\mu_{\bm r}}$ and
the spin-current tensor is given by the Eqs.(\ref{SC4}) and (\ref{SC5})
with $\Omega\to\Omega(\mu_{\bm r})$ and $D\to D(\mu_{\bm r})=
D|_{E=\mu_{\bm r}}$. In Eq.(\ref{F9}) we have also included the 
variables the densities depend on   
to stress that the densities are functions of 
the total  energy $E$, of the particle 
position ${\bm r}$ and of $t$ in the presence of a field.
Below we omit the variables again to avoid cluttering the notation.

At this point it is important to note the following:
whereas Eq.(\ref{SD3}) can only be written in the form (\ref{SC9})
due to the identity (\ref{SC8}), Eq.(\ref{F9}) contains also 
derivatives of spin-charge coupling coefficients, since the
equality (\ref{SC8}) does not hold in the presence of electric fields.
The spin-charge coupling term (second term on the rhs of Eq.(\ref{SC5}))
arises naturally in the diffusion equation. It does not reduce to the
divergence of a curl. Consequently, we can truly claim that the spin-charge
coupling terms enter the diffusion equation and thus, that the spin-current
tensor coincides with the tensor of transport currents. The second point we 
would like to draw attention to is the following:
the spin-charge coupling terms would be absent
if we had ignored the 
distortion of the particle spectrum due to the electric field.

Eq.(\ref{F9}) couples the spin density with the charge density. The diffusion
equation for the charge density takes the form
\begin{equation}\label{F10}
\partial_t n({\bm r}|E,t)+\mbox{div}{\bm j}({\bm r}|E,t),
\end{equation}
where
\begin{equation}\label{F11}
{\bm j}({\bm r}|E,t)=-D(\mu_{\bm r}){\bm\nabla}n({\bm r}|E,t)
-\frac{3\hbar}{2}\frac{\tau}{m}\Omega(\mu_{\bm r}){\bm e}_z\times{\bm\nabla}
S_z({\bm r}|
E,t).
\end{equation}
Eq.(\ref{F11}) sets the situation again apart from that in the
linear Rashba-model. Homogeneous
non-equilibrium magnetizations in the plane of the 2-d electron gas
lead to electric currents in the linear Rashba-model
in the absence of external fields (the spin-galvanic currents). 
In the cubic Rashba model, by contrast,
non-equilibrium magnetizations can only induce electric
currents in the 
absence of external fields if they 
are inhomogeneous and perpendicular to the 2-d plane
of the 2-d electron gas. A homogeneous non-equilibrium
magnetization perpendicular to the 2-d plane can only induce a current in 
the presence of a field, as discussed further below. 

The Eqs.(\ref{F9})-(\ref{F11}) do not have the structure of conventional
diffusion equations yet. The reason for this fact is that these equations
are written down for densities, which depend on the total energy of the 
particles. To obtain conventional diffusion equations we
distinguish between the position dependence of the densities, 
which characterizes the shape of the packet, and the position dependence of the
reference point of the kinetic energy, viz we write the densities
in the form $n({\bm r}|E,t)\to n({\bm r}|\mu_{\bm r},t)$,
${\bm S}({\bm r}|E,t)\to {\bm S}({\bm r}|\mu_{\bm r},t)$ and replace
$\mu_{\bm r}$ by $\epsilon$. The position dependence
of the density $n({\bm r}|\epsilon,t)$ describes
the shape of a particle packet of particles with the same kinetic energy
$\epsilon$,
which is measured from the tilted bottom of the band. Likewise
for the density ${\bm S}({\bm r}|\epsilon,t)$. Doing so, we find that the
derivatives transform according to the rule ${\bm \nabla}\to
{\bm\nabla}-{\bm F}\partial_{\epsilon}$. Using this method, we obtain
\begin{equation}\label{F12}
\partial_t {S}_k+{\bm\Omega}_{ki}(\epsilon){S}_i+\nabla_iJ_{ik}=-
\partial_{\epsilon}{j_{\epsilon}}_i,
\end{equation}
where ${j_{\epsilon}}_i$ is the energy relaxation current for spins,
\begin{equation}\label{F13}
J_{ix}=-D(\epsilon)\nabla_iS_x +D(\epsilon)F_i\partial_{\epsilon}S_x,
\hspace{1ex}J_{iy}=-D\nabla_iS_y+D(\epsilon)F_i\partial_{\epsilon}S_y,
\end{equation}
and
\begin{equation}\label{F14}
J_{iz}=-D(\epsilon)\nabla_iS_z+D(\epsilon)F_i\partial_{\epsilon}S_z
+\frac{3}{2}\frac{\hbar}{m}\tau\Omega(\epsilon)
({\bm e}_z\times{\bm\nabla })_in
-\frac{3}{2}\frac{\hbar}{m}\tau\Omega(\epsilon)
({\bm e}_z\times{\bm F})_i\partial_{\epsilon}n
.
\end{equation}
The second term on the rhs of the Eqs.(\ref{F13}) and (\ref{F14})
might look unfamiliar. However, in looking on these equations one has to 
take into account that these equations are spectral diffusion equations.
To obtain diffusion equations, say for particles at the Fermi surface,
one has to integrate these equations with respect to $\epsilon$. In this case
the first term on the rhs of Eq.(\ref{F12}) vanishes and the second term
on the rhs of the Eqs.(\ref{F13}) and (\ref{F14}) yields just the mobility 
times the field, as expected.

The electric current takes the following form
\begin{equation}\label{F15}
{\bm j}=-D(\epsilon){\bm\nabla}n+D(\epsilon){\bm F}\partial_{\epsilon}n
-\frac{3\hbar}{2}\frac{\tau}{m}\Omega(\epsilon){\bm e}_z\times{\bm\nabla}
S_z+\frac{3\hbar}{2}\frac{\tau}{m}\Omega(\epsilon){\bm e}_z\times{\bm F}
\partial_{\epsilon} S_z.
\end{equation}
and the diffusion equation (\ref{F10}) yields
\begin{equation}\label{F17}
\partial_t n+\mbox{div}{\bm j}=-\partial_{\epsilon}j_{\epsilon},
\end{equation}
where $j_{\epsilon}$ is the energy relaxation current for particles.  
Note that, the fourth term on the rhs of Eq.(\ref{F15}) yields also
an electric current in an electric field if there is 
a homogeneous magnetization in 
$z$-direction. This current is a Hall current, i.e., it is transverse
to ${\bm F}$.
\subsection{Spin-Hall effect}
To investigate the spin-Hall effect we focus on a system
in the half-plane $y>0$.  We assume that the system is 
translation invariant in $x$-direction ($d/dx=0$), that the electric field
is applied in $x$-direction and that
there is an additional scattering process, which keeps the system
in energy equilibrium. Thus, $n({\bm r}|\epsilon,t)=n_0
\theta(\epsilon_F-\epsilon)$, where $\epsilon_F$ is the Fermi-energy
and $n_0=2\nu$. A glance on the spin diffusion equations (\ref{F12})-
(\ref{F14}) reveals that a non-equilibrium magnetization at the Fermi 
surface is created in this case.
Therefore, we can solve the spin diffusion equation with the ansatz
${\bm S}({\bm r}|\epsilon,t)={\bm S}({\bm r}|t)\delta(\epsilon-\epsilon_F)$.
To obtain an equation for ${\bm S}({\bm r}|t)$ we restrict the
consideration to the stationary limit and integrate the spin
diffusion equations with respect to $\epsilon$. Doing so, we obtain the
equation
\begin{equation}\label{SH1}
2\Omega(\epsilon_F)S_z-D(\epsilon_F)\frac{d^2}{dy^2}S_z=0,
\end{equation}
which has to be solved subjected to the boundary conditions
\begin{equation}\label{SH2}
0=J_{yz}|_{y=0}=-D(\epsilon_F)\frac{d}{dy}S_z+
\frac{3\hbar}{2m}\Omega(\epsilon_F)\tau F_x n_0
\end{equation}
and $S_z=$ finite for $y\to \infty$. The unique solution is
\begin{equation}\label{SH3}
S_z(y)=-\frac{3}{2}\sqrt{\frac{\Omega(\epsilon_F)}{2D(\epsilon_F)}}
\frac{\hbar}{m}\tau
n_0F_x\exp(-\sqrt{2\Omega(\epsilon_F)/D(\epsilon_F)}y).
\end{equation}
Thus, the spin-Hall current in the bulk leads to a 
non-equilibrium magnetization at the boundary, as it is
also the case in systems with strong Rashba spin-orbit
interaction\cite{Nomura}.  Since the exponent
depends only on the ratio $\Omega(\epsilon_F)/D(\epsilon_F)$
the penetration depth of the spin accumulation is independent
of disorder in systems with weak spin-orbit interaction. It
decreases with increasing particle number. The magnitude
of the edge spin-accumulation increases with increasing $\epsilon_F$ since
$\Omega(\epsilon_F)\propto\epsilon_F^3$ .
This sets
the situation apart from that in nearly clean systems. There 
both the magnitude of the edge spin-accumulation and the penetration depth
decrease with increasing particle number\cite{Nomura}.

The 
disorder enters only in the preexponential factor via the relaxation time. 
This factor leads to a reduction of the magnitude of the 
magnetization with increasing amount of disorder.
\subsection{Hall-like current}
Spin-charge coupling effects manifest themselves also in the 
electric current in optical experiments. To investigate the magnitude
of the Hall-like current, we assume that a constant non-equilibrium 
magnetization 
of the type ${\bm S}({\bm r}|\epsilon,t)=S_z{\bm e}_z
\delta(\epsilon-\epsilon_F)$ is created optically
by means of optical orientation. Such a density leads to a Hall-like
current of the form
\begin{equation}\label{HC1}
\delta{\bm j}=-\frac{3}{2}\frac{\hbar\tau}{m}{\bm e}_z\times{\bm F} S_z
\frac{d\Omega(\epsilon_F)}{d\epsilon_F}.
\end{equation}
Surprisingly  this current depends only quadratically on $N$, as it is 
also the case for the spin-Hall effect. It shares this property with the
Hall current, which has recently been discussed in Ref.[\onlinecite{Niu}].
We speculate that the physics is similar.  
\section{Conclusions}
We have studied the diffusion equations for heavy holes in semiconductor 
quantum wells with Rashba spin-orbit interaction. The holes  are 
described by a cubic Rashba-model. In our investigation we have restricted the
consideration to systems with weak Rashba interaction. These systems have the 
property that the energy level splitting $\Delta$ due to the Rashba 
interaction at the Fermi energy is small compared to the disorder energy 
$\hbar/2\tau$. 

At first glance the cubic Rashba model differs from the conventional 
linear Rashba model  
only in that the interaction does not depend  linearly
but cubically on the momentum. This small difference, however,
leads to drastic changes in the structure of the spin diffusion equations. 
The particle and spin diffusion equations 
reduce to conventional diffusion equations in the absence of
external fields. There
 are no spin-charge coupling effects in the diffusion equations themselves
in the absence of external fields. There is neither a spin-accumulation,
which can be induced by concentration gradient, nor a spin-galvanic current
in homogeneous non-equilibrium systems. This sets the situation strongly
apart from that for systems with linear Rashba-interaction, in which 
spin-charge coupling terms occur in the particle and spin diffusion equations,
which are related to the existence of these phenomena. If we switch
on an external field the situation changes. Spin-charge coupling
terms occur also explicitly in the diffusion equation, which lead
to a non-equilibrium current in the presence of a perpendicular
non-equilibrium magnetization and to spin-Hall effect. Our investigation
shows that both effects depend only on the magnitude of the
Rashba interaction strength. Explicit
formulas for the magnitude of the spin-accumulation at the boundary
and for the magnitude of the non-equilibrium current under steady
state illumination are given in the text.

The investigation of the structure of the spin-diffusion equations
has also yield further informations on the physics of spin-diffusion processes
in such systems. It has shown that the spin-current tensor coincides with 
the tensor of transport currents in the diffusion equation (see 
Eq.(\ref{SC8a}) for the definition of tensor of transport currents). 
Therefore, the spin-current tensor determines the boundary conditions 
needed to find solutions to the spin diffusion equations. Due to this fact 
spin-charge coupling effects enter the solution to the diffusion 
equations even in the absence of external fields, although they are
absent in the diffusion equations themselves in the absence of 
fields. This fact is remarkable and sets the situation again strongly
apart from that in the linear model. In the linear model there is no
relationship between the tensor of transport currents and the 
spin-current tensor\cite{OB2}. 

The huge differences between the results for the linear and the cubic
model raise the question, why the physics in the cubic Rashba model 
differs so strongly from that in the linear model. We have no clear
answer to  the question yet but speculate that the difference is caused 
by the fact that the linear spin-orbit coupling has nearly the structure 
of a pure gauge. In fact, to take into account the spin-orbit interaction
in the linear model  we only
have to replace the derivatives by covariant derivatives in the diffusion 
equations, as long as 
spin-charge coupling effects are ignored\cite{OB5}.  

\begin{acknowledgments}
The authors are grateful to H. B\"ottger and P. Kleinert for many useful and
interesting discussions on the subject.
\end{acknowledgments}


\begin{thebibliography}{50}
\bibitem{Sinova} J. Sinova, D. Culcer, Q. Niu, N. A. Sinitsyn, T. Jungwirth, 
and A. H. MacDonald, Phys. Rev. Lett. {\bf 92}, 126603 (2004). 
\bibitem{Wunderlich} J. Wunderlich, B. Kaestner, J. Sinova, 
T. Jungwirth, Phys. Rev. Lett. {\bf 94}, 047204 (2005).
\bibitem{Nikolic} B. K. Nikoli{\'c}, L. P. Z{\^a}bo, and S. Souma, Phys. Rev. B {\bf 72}, 075361 (2005), Phys. Rev. B {\bf 73}, 075303 (2006).
\bibitem{Nikolic2} B. K. Nikoli{\'c} and S. Souma, Phys. Rev. B {\bf 71}, 195328 (2005).
\bibitem{Loss} J. Schliemann and D. Loss, Phys. Rev. B {\bf 69}, 165315 (2004).
\bibitem{Sheng} L. Sheng, D. N. Sheng, and C. S. Ting, Phys. Rev. Lett. {\bf 94}, 016602 (2005).
\bibitem{Murakami} S. Murakami, N. Nagaosa, and S. C. Zhang, Science
{\bf 301}, 1348 (2003) (see also the supporting online material).
\bibitem{Bernevig} B. A. Bernevig, and S. C. Zhang , Phys. Rev. Lett. {\bf 95}, 016801 (2005).
\bibitem{Nomura} K. Nomura, J. Wunderlich, J. Sinova, B. Kaestner,
A. H. MacDonald, and T. Jungwirth, Phys. Rev. B {\bf 72}, 245330 (2005).
\bibitem{Nomura2} K. Nomura, J. Sinova, N. A. Sinitsyn, and A. H. MacDonald,
Phys. Rev. B {\bf 72}, 165316 (2005).
\bibitem{Wu} M. W. Wu and J. Zhou, Phys. Rev. B {\bf 72}, 115333 (2005).
\bibitem{Kato} Y. K. Kato, R. C. Myers, A. C. Gossard, and D. D. Awschalom, 
Phys. Rev. Lett. {\bf 93}, 176601 (2004).
\bibitem{Silov} A. Yu. Silov, P. A. Blajov, J. H. Woller, R. Hey, K. H. Ploog, and N. S. Averkiev, Appl. Phys. Lett. {\bf 85}, 5929 (2004).
\bibitem{Ganichev} S. D. Ganichev, S. N. Danilov, P. Schneider, V. V. Bel'kov, 
L. E. Golub, W. Wegschneider, D. Weiss, and W. Prettl, cond-mat/0403641
(unpublished).
\bibitem{Edelstein} V. M. Edelstein, Solid State Commun. {\bf 73}, 233 (1990).
\bibitem{Inoue} J. I. Inoue, G. E. W. Bauer, and L. W. Molenkamp,
Phys. Rev. B {\bf 67}, 033104 (2003).
\bibitem{Culcer} D. Culcer, Y. Yao, A. H. MacDonald, and Q. Niu, 
Phys. Rev. B {\bf 72}, 045215 (2005).
\bibitem{Kleinert2} P. Kleinert, V. V. Bryksin and O. Bleibaum, 
Phys. Rev. B {\bf 72}, 195311 (2005).
\bibitem{Raimondi} R. Raimondi, P. Schwab, Phys. Rev. B {\bf 71}, 33311 (2005).
\bibitem{Chalaev}  O. Chalaev and D. Loss, Phys. Rev. B {\bf 71}, 
245318 (2005).
\bibitem{Murakami1} S. Murakami, Phys. Rev. B {\bf 69}, 241202(R) (2004).
\bibitem{Rashba} E. I. Rashba, Phys. Rev. B {\bf 68}, 241315(R)
(2003), J. Supercond.{\bf 18}, 137 (2005).
\bibitem{Zhang} S. Zhang and Z. Yang, Phys. Rev. Lett. {\bf 94}, 066602 
(2005).
\bibitem{Burkov} A. A. Burkov, A. S. Nunez and A. H. MacDonald, 
Phys. Rev. B {\bf 70}, 155308 (2004).
\bibitem{Mishchenko} E. G. Mishchenko, A. V. Shytov, and B. I. Halperin, 
Phys. Rev. Lett. {\bf 93}, 226602 (2004).
\bibitem{OB1} O. Bleibaum, Phys. Rev. B {\bf 71}, 195329 (2005).
\bibitem{OB2} O. Bleibaum, Phys. Rev. B {\bf 73}, 35322 (2006).
\bibitem{Bauer} I. Adagideli, G. E. W. Bauer, Phys. Rev. Lett. {\bf 95}, 
256602 (2005).
\bibitem{OB4} O. Bleibaum, Phys. Rev. B {\bf 72}, 75366 (2005).
\bibitem{Damker} T. Damker, H. B\"ottger and V. V. Bryksin, Phys. Rev. B 
{\bf 69} 205327 (2004).
\bibitem{OB5} O. Bleibaum, cond-mat/0503471 (unpublished) (2005).
\bibitem{Zhang2} P. Zhang, J. Shi, Si Xiao, and Q. Niu, cond-mat/0503505v4 
(2005).
\bibitem{Sugimoto} N. Sugimoto, S. Onoda, S. Murakami and N. Nagaosa, 
cond-mat/0503475v3 (2006). 
\bibitem{Krotkov} P. L. Krotkov, and S. Das Sarma, cond-mat/0510114v1 (2005).
\bibitem{Schliemann} J. Schliemann and D. Loss, Phys. Rev. B {\bf 71}, 85308
(2005). 
\bibitem{Zulicke1} M. G. Pala, M. Governale, J. K\"onig, U. Z\"ulicke,
G. Iannaccone, Phys. Rev. B {\bf 69}, 045304 (2004).
\bibitem{Zulicke2} M. G. Pala, M. Governale, J. K\"onig, U. Z\"ulicke,
Europhys. Lett. {\bf 65}, 850 (2004).
\bibitem{Winkler1} R. Winkler, Phys. Rev. B {\bf 62}, R4245 (2000).
\bibitem{Winkler2} R. Winkler, H. Noh, E. Tutuc, and M. Shayegan, Phys. Rev.
B {\bf 65}, 155303 (2002).
\bibitem{OB3} O. Bleibaum, Phys. Rev. B {\bf 69}, 205202 (2004).
\bibitem{Wu1} M. Q. Weng, M. W. Wu, and L. Jiang, Phys. Rev. B {\bf 69}, 
245320 (2004). 
\bibitem{Awschalom1} Y. K. Kato, R. C. Myers, A. C. Gossard, and 
D. D. Awschalom, Appl. Phys. Lett. {\bf 87}, 22503 (2005). 
\bibitem{Kl} P. Kleinert and V. V. Bryksin, Phys. Rev. B (in print).
\bibitem{Landau} L. D. Landau and E. M. Lifschitz, 
{\it Lehrbuch der Theoretischen Physik VI Hydrodynamik} (Akademie-Verlag, Berlin 1991).
\bibitem{remark} Note that, since ${\bm S}$ is a vector we could as well
expect that the diffusion equation can be written in the form
\[\partial_tS_k+{\bm\Omega}_{ki}S_i+\nabla_i{\tilde J}_{ik}+(\mbox{curl}T)_k
=0,\] where $T$ is a vector field, which depends on $S_{\gamma}$. The choice
of the structure (\ref{SC9}) is finally only based on the observation, 
that  hydrodynamic equations can usually be written in the form
(\ref{SC9}) (see, e.g., [\onlinecite{Landau}]).
\bibitem{Niu} P. Zhang and Q. Niu, cond-mat/0406436v1 (2004).
%
\end{thebibliography}
\end{document}